\documentclass{article}
\usepackage{amsmath,amsthm}
\usepackage{amssymb,latexsym}
\usepackage[mathscr]{eucal}
\usepackage{setspace}
\usepackage{graphics}
\graphicspath{{converted_graphics/}}
\usepackage{epsf}
\usepackage{graphicx}
\usepackage{array}

\newcommand{\pa}{\partial}

\newcommand{\om}{\omega}

\newcommand{\ga}{\gamma}

\newcommand{\lp}{\left(}
\newcommand{\rp}{\right)}
\newcommand{\lb}{\left[}
\newcommand{\rb}{\right]}
\newcommand{\lc}{\left\{}
\newcommand{\rc}{\right\}}

\newcommand{\beq}{\begin{equation}}
\newcommand{\eq}{\end{equation}}

\newcommand{\tex}{\textit}
\newcommand{\no}{\noindent}

\begin{document}

\title{\bf Resonant Mode Conversion of the Second Sound to the First Sound in Liquid Helium II}        % Enter your title between curly braces
\author{B. K. Shivamoggi\\
University of Central Florida\\
Orlando, FL 32816-1364, USA\\
}        % Enter your name between curly braces
\date{}          % Enter your date or \today between curly braces
\maketitle

\noindent \large{\bf Abstract} \\ \\

The experimentally observed disappearance below $T=0.5 K$ of the second sound in liquid $He$ II as a separate wave mode and its subsequent propagation at the speed of the first sound (Peshkov \cite{pes}) may be interpreted as a resonant mode conversion of the second sound to the first sound. Near the resonant mode coupling point $T=T_\ast$, where the frequencies of the two waves become equal, the anomalous effect of entropy changes on the first sound and density changes on the second sound, though generally small, become significant. This leads to the resonant mode coupling of the first sound and the second sound and forces them to lose their identities and hence pave the way for the resonant mode conversion of the second sound to the first sound. We give a theoretical framework for this proposition and an estimate for the fraction of the second sound that is mode-converted to the first sound.

\pagebreak

\noindent\large{\bf 1. Introduction}\\

In addition to the pressure waves of ordinary sound, in which the normal fluid and superfluid components in liquid $He$ II move in phase, Tisza \cite{tis} predicted the existence of a new wave, called the "\tex{second sound}"\footnote{The word "\tex{sound}" in "\tex{second sound}" is rather misleading (London \cite{lon}) because these waves are not pressure waves like ordinary sound and could not be excited with a piezo-electric crystal. In the second sound, heat is transferred via a quantum wave-like motion rather than via the classical diffusion.}, in which the normal fluid and superfluid components move perfectly out of phase. The second sound is a compression wave of the entropy with the mass density remaining nearly constant. The existence of second sound in liquid $He$ II was experimentally confirmed by Peshkov \cite{pes} who also determined the temperature dependence of its propagation speed. The second sound has served as a very valuable tool in superfluid dynamics research - the of second sound was used as a diagnostic (Hall and Vinen \cite{hal}) to study the \textit{mutual friction} (Feynman \cite{fey}) underlying the interaction between the normal fluid and superfluid. 

\vspace{.10in}

Below $T\approx 0.5 K$, the normal fluid component almost disappears (Andronikashvili \cite{and}) thereby impairing the establishment of the thernodynamic equilibrium and the second sound ceases to exist as a separate wave mode and propagates at the speed of the first sound (while the speed of the first sound is hardly affected by temperature (Peshkov \cite{pes}, McClintock et al. \cite{mcc})). The purpose of this paper is to propose that this process may be interpreted as a resonant mode conversion of the second sound to the first sound\footnote{Onsager (Donnelly \cite{don}) interestingly put forward the idea of converting second sound to first sound at a liquid-vapor surface as a means of detecting second sound in experiments, which was indeed used by Lane et al. \cite{la}.}. We give a theoretical framework for this proposition which is based on the resonant coupling between the second sound and the first sound.

\vspace{.25in}

\noindent\large{\bf 2. Resonant Mode Coupling of the First Sound and the Second Sound}\\

In order to discuss resonant mode conversion process in question, we need to consider the sound propagation in a variable temperature liquid $He$ II. For the moment, however, we consider sound propagation in a constant temperature liquid $He$ II so that the dispersion relation considered below may be viewed as a \textit{local} relation in the temperature space.

Looking for plane-wave solutions of the form,

\beq%1
q\lp x,t\rp\sim e^{i\lp \om t-kx\rp}
\eq

\no the linearized two-fluid equations for liquid $He$ II lead to the dispersion relation (Landau \cite{lan}), 

\beq%2
\lp\om^2-\om^2_1\rp \lp\om^2-\om^2_2\rp = \om^2_1\om^2_2\left(\frac{\ga-1}{\ga}\right)
\eq

\no where (in usual notation),

\beq\notag
\begin{matrix}
\begin{aligned}
\om_1&\equiv ku_1,~\om_2=ku_2\\
u_1&\equiv\left(\frac{\pa p}{\pa \rho}\right)_S, ~u_2\equiv\frac{\rho_S}{\rho_n}S^2\left(\frac{\pa T}{\pa S}\right)_\rho\\
\ga&\equiv\frac{C_p}{C_v }
\end{aligned}
\end{matrix}
\eq

\vspace{.10in}

Equation (2) describes the coupling of the second sound (denoted by subscript 2) with the first sound (denoted by subscript 1).

We now allow the temperature of liquid $He$ II to vary; the normal-fluid density $\rho_n$ and the superfluid density $\rho_S,$ being functions of temperature, now become variables. In general, entropy changes do not affect the first sound while the density changes do not affect the second sound, which is expressed by the result that $\gamma\approx 1$ for liquid $He$ II. Consequently, as per equation (2), the first sound and the second sound generally propagate almost independently, and their frequencies are given by 

\beq%3
\om_1=\om_1\lp k,T\rp, \om_2=\om_2\lp k, T\rp.
\eq

\no However, near the resonant mode-coupling point $T=T_\ast$\footnote{$T_\ast$ is between 0 and 0.5K. The mean free path in liquid $He$ II is found to go up at these low temperatures, so rigorously speaking, the fluid model leading to equation (2) can become suspect.}, where the frequencies of the two waves become equal,

\beq%4
\om_1\lp k_\ast, T_\ast\rp=\om_2\lp k_\ast, T_\ast\rp
\eq

\no the anomalous effect of entropy changes on the first sound and density changes on the second sound, though generally small, become significant. This leads to the resonant mode coupling of the first sound and the second sound - mathematically, the coupling constant in equation (2) materializes near $T=T_\ast$, the closeness of $\gamma$ to $1$ notwithstanding. The resonant mode coupling forces the two waves to lose their identities. 

Near $T=T_\ast$, equation (2) may be approximated by

\beq%5
\lp\om-\om_1\rp \lp\om-\om_2\rp\approx\frac{1}{4} k^2_\ast u^2_{1_\ast}\left(\frac{\ga-1}{\ga}\right).
\eq

\begin{figure}
\begin{center}
\includegraphics[width=7.2in,angle=00]{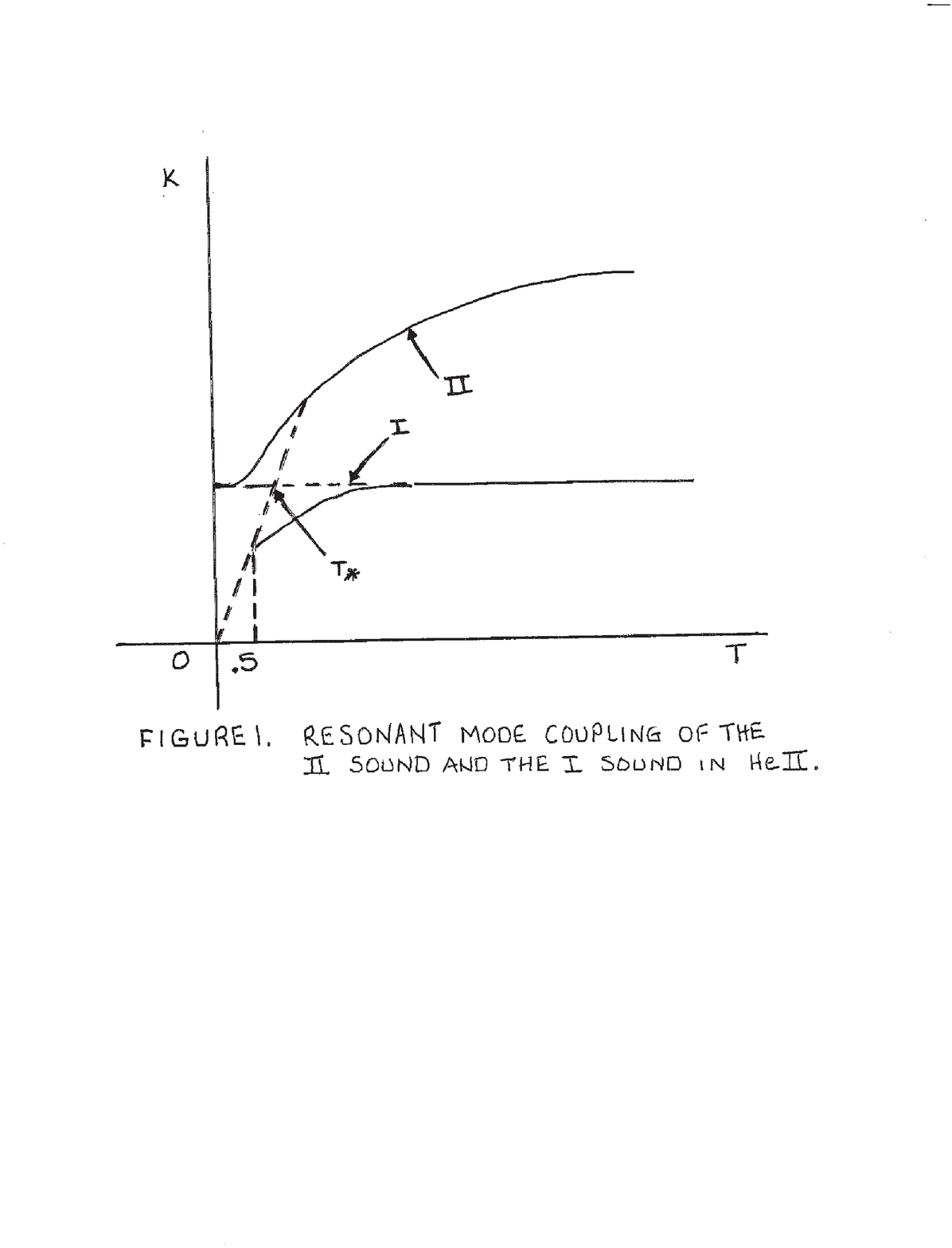}{}
\label{Figure 1.}
\end{center}
\end{figure}
\no As a result of this resonant mode coupling, the second sound gets mode-converted to the first sound.\footnote{If the second sound is launched by an external driver, then the mode conversion to first sound may be expected to depend on how it interacts with the boundary conditions compatible with the driver.} The uninterrupted lines in Figure 1 show the way in which the two waves couple near $T=T_\ast$ where the anomalous effects of the entropy changes and the density changes on the two waves materialize. Figure 1 shows that if the second sound is under control in an experiment, then it would change its temperature dependence near $T=T_\ast$.

Equation (5) is of the same form as the model equation for resonant mode conversion of plasma waves considered by Cairns and Lashmore-Davies \cite{cai}. Using the results in \cite{cai}, the fraction of the second sound that is mode-converted to the first sound is given by 

\beq%6
C=1-\exp\lc-\lb\frac{\pi k^2_\ast u^2_{1_\ast}\left(\dfrac{\ga-1}{\ga}\right)}{u_{2_\ast}\left(\pa\om_1/\pa T\right)_\ast-u_{1_\ast}\left(\pa\om_2/\pa T\right)_\ast}\rb\rc.
\eq

\vspace{.10in}

\noindent\large{\bf 3. Discussion}\\
	
In this paper, the experimentally observed disappearance below $T=0.5 K$ of the second sound in liquid $He$ II as a separate wave mode and its subsequent propagation at the speed of the first sound (Peshkov \cite{pes}) has been proposed to be viewed as a resonant mode conversion of the second sound to the first sound. Near the resonant mode-coupling point $T=T_\ast$, where the frequencies of the two waves become equal, the anomalous effect of entropy changes on the first sound and density changes on the second sound, though generally small, become significant. This leads to the resonant mode coupling of the first sound and the second sound and forces them to lose their identities and hence pave the way for the resonant mode conversion of the second sound to the first sound. We have given a theoretical framework for this proposition and an estimate for the fraction of the second sound that is mode-converted to the first sound.  

\vspace{.10in}

\noindent\large\textbf{Acknowledgements}\\

\vspace{.10in}

This work was initiated during the course of my participation in the \textit{Vorticity Program} at the Aspen Center for Physics, which is supported in part by the National Science Foundation. I am very thankful to Grisha Falkovich, Ladik Skrbek and Peter Weichman for their valuable remarks.

\end{document}